\documentclass{svmult}

\usepackage{makeidx}         % allows index generation
\usepackage{graphicx}        % standard LaTeX graphics tool
\usepackage{multicol}        % used for the two-column index
\usepackage[bottom]{footmisc}% places footnotes at page bottom
\usepackage{amssymb}

\makeindex             % used for the subject index
\begin{document}

\title*{Notes on Canonical Quantization of Symplectic Vector Spaces over Finite Fields}
\titlerunning{Notes on Quantization of Symplectic Vector Spaces}
\author{Shamgar Gurevich$^*$ and Ronny Hadani$\dagger$}
\institute{$^*$ Department of Mathematics, University of California, Berkeley, CA
94720, USA
\texttt{shamgar@math.berkeley.edu} \\
$\dagger$ Department of Mathematics, University of Chicago, IL 60637, USA
\texttt{hadani@math.uchicago.edu}}
\authorrunning{S. Gurevich and R. Hadani}
\maketitle

\begin{abstract}
In these notes we construct a quantization functor, associating a Hilbert
space $\mathcal{H}(V)$ to a finite dimensional symplectic vector space $V$
over a finite field $\mathbb{F}_{q}$. As a result, we obtain a canonical
model for the Weil representation of the symplectic group $Sp\left( V\right)
$. The main technical result is a proof of a stronger form of the Stone-von
Neumann theorem for the Heisenberg group over $\mathbb{F}_{q}$. Our result
answers, for the case of the Heisenberg group, a question of Kazhdan about
the possible existence of a canonical Hilbert space attached to a coadjoint
orbit of a general unipotent group over $\mathbb{F}_{q}$.
\end{abstract}

\keywords{Quantization functor, Weil representation, Quantization of Lagrangian correspondences, Geometric intertwining operators.}
\\
\\
\textbf{AMS codes:} 11F27, 53D50.

%%-----------------------------------------------------------------------
\section{Introduction}

Quantization is a fundamental procedure in mathematics and in physics.
Although it is widely used in both contexts, its precise nature remains to
some extent unclear. From the physical side, quantization is the procedure
by which one associates to a classical mechanical system its quantum
counterpart. From the mathematical side, it seems that quantization is a way
to construct interesting Hilbert spaces out of symplectic manifolds,
suggesting a method for constructing representations of the corresponding
groups of symplectomorphisms \cite{Ki62, Ko70}.

Probably, one of the principal manifestation of quantization in mathematics
appears in the form of the Weil representation \cite{S63, Sh62, W64} of the
metaplectic group
\begin{eqnarray}
\rho :Mp\left( 2n,%
%TCIMACRO{\U{211d} }%
%BeginExpansion
\mathbb{R}
%EndExpansion
\right) \rightarrow U\left( L^{2}\left(
%TCIMACRO{\U{211d} }%
%BeginExpansion
\mathbb{R}
%EndExpansion
^{n}\right) \right) , \nonumber
\end{eqnarray}%
where $Mp\left( 2n,%
%TCIMACRO{\U{211d} }%
%BeginExpansion
\mathbb{R}
%EndExpansion
\right) $ is a double cover of the linear symplectic group $Sp\left( 2n,%
%TCIMACRO{\U{211d} }%
%BeginExpansion
\mathbb{R}
%EndExpansion
\right) $. The general ideology \cite{We81, We82} suggests that the Weil
representation appears through a quantization of the standard symplectic
vector space $\left(
%TCIMACRO{\U{211d} }%
%BeginExpansion
\mathbb{R}
%EndExpansion
^{2n},\omega \right) $. This means that there should exist a quantization
functor $\mathcal{H}$, associating to a symplectic manifold $\left( M,\omega
\right) $ an Hilbert space $\mathcal{H}\left( M\right) $, such that when
applied to $\left(
%TCIMACRO{\U{211d} }%
%BeginExpansion
\mathbb{R}
%EndExpansion
^{2n},\omega \right) $ it yields the Weil representation in the form of
\begin{eqnarray}
\mathcal{H}:Sp\left( 2n,%
%TCIMACRO{\U{211d} }%
%BeginExpansion
\mathbb{R}
%EndExpansion
\right) \rightarrow U\left( \mathcal{H}\left(
%TCIMACRO{\U{211d} }%
%BeginExpansion
\mathbb{R}
%EndExpansion
^{2n},\omega \right) \right). \nonumber
\end{eqnarray}%
As stated, this ideology is too naive since it does not account for the
metaplectic cover.

\subsection{Main results}

In these notes, we show that the quantization ideology can be made precise
when applied in the setting of symplectic vector spaces over the finite
field $\mathbb{F}_{q}$, where $q$ is odd. Specifically,
we construct a quantization functor $\mathcal{H}:$ $\mathsf{Symp}\rightarrow
\mathsf{Hilb}$, where $\mathsf{Symp}$ denotes the (groupoid) category whose
objects are finite dimensional symplectic vector spaces over $\mathbb{F}_{q}$
and morphisms are linear isomorphisms of symplectic vector spaces and $%
\mathsf{Hilb}$ denotes the category of finite dimensional Hilbert spaces.

As a consequence, for a fixed symplectic vector space $V\in $ $\mathsf{Symp}$%
, we obtain, by functoriality, a homomorphism $\mathcal{H}:Sp\left( V\right)
\rightarrow U\left( \mathcal{H}\left( V\right) \right) $, which we refer to
as the canonical model of the Weil representation of the symplectic group $%
Sp\left( V\right) $.

\subsubsection{Properties of the quantization functor}

In addition, we show that the functor $\mathcal{H}$ satisfies the following
basic properties (cf. \cite{We82}):

\begin{itemize}
\item \textbf{Compatibility with Cartesian products.} The functor $\mathcal{H%
}$ is a monoidal functor: Given $V_{1},V_{2}\in \mathsf{Symp}$, we have a
natural isomorphism
\begin{eqnarray}
\mathcal{H}\left( V_{1}\times V_{2}\right) \simeq \mathcal{H}\left(
V_{1}\right) \otimes \mathcal{H}\left( V_{2}\right) . \nonumber
\end{eqnarray}

\item \textbf{Compatibility with duality. }Given $V=\left( V,\omega \right)
\in \mathsf{Symp}$, its symplectic dual is $\overline{V}=\left( V,-\omega
\right) $. There exists a natural non-degenerate pairing
\begin{eqnarray}
\left \langle \mathcal{\cdot },\mathcal{\cdot }\right \rangle _{V}:\mathcal{H%
}\left( \overline{V}\right) \times \mathcal{H}\left( V\right) \rightarrow
%TCIMACRO{\U{2102} }%
%BeginExpansion
\mathbb{C}
%EndExpansion
.    \nonumber
\end{eqnarray}

\item \textbf{Compatibility with linear symplectic reduction}. Given $V\in
\mathsf{Symp}$, $I\subset V$ an isotropic subspace in $V$ and $o_{I}\in
\wedge ^{\textrm{top}}I$ \ a non-zero vector, there exists
a natural isomorphism
\begin{equation}
\mathcal{H}\left( V\right) ^{I}\simeq \mathcal{H}\left( I^{\bot }/I\right) ,
\label{reduc_eq}
\end{equation}%
where $\mathcal{H}\left( V\right) ^{I}$ stands for the subspace of $I$%
-invariant vectors in $\mathcal{H}\left( V\right) $ \ (an operation which
will be made precise in the sequel) and $I^{\perp }/I\in \mathsf{Symp}$ is
the symplectic reduction of $V$ with respect to $I$ \cite{BW97}. (A pair $\left( I,o_{I}\right) $, where $o_{I}\in \wedge ^{\textrm{top}}I$ is non-zero vector, is called an \textit{oriented isotropic subspace}).
\end{itemize}

\textbf{Quantization of oriented Lagrangian subspaces}. A particular
situation is when $I=L$ is a Lagrangian subspace. In this situation, $%
L^{\perp }/L=0$ and (\ref{reduc_eq}) yields an isomorphism $\mathcal{H}%
\left( V\right) ^{L}\simeq \mathcal{H}\left( 0\right) =%
%TCIMACRO{\U{2102} }%
%BeginExpansion
\mathbb{C}
%EndExpansion
$, which associates to $1\in
%TCIMACRO{\U{2102} }%
%BeginExpansion
\mathbb{C}
%EndExpansion
$ a vector $v_{L^{\circ }}\in \mathcal{H}\left( V\right) $. This means that
we establish a mechanism which associates to every oriented Lagrangian
subspace in $V$ a well defined vector in $\mathcal{H}(V)$%
\begin{eqnarray}
L^{\circ }\longmapsto v_{L^{\circ }}\in \mathcal{H}(V). \nonumber
\end{eqnarray}%
Interestingly, to the best of our knowledge (cf. \cite{GS79}), this kind of
structure, which exists in the setting of the Weil representation of the
group $Sp\left( V\right) ,$ was not observed before.

\textbf{Quantization of oriented Lagrangian correspondences.} It is also
interesting to consider simultaneously the compatibility of $\mathcal{H}$
with Cartesian product, duality, and linear symplectic reduction. The first
and second properties imply that $\mathcal{H}\left(\overline{V}_{1}\times V_{2}\right) $ is naturally isomorphic to the vector space $\mathrm{Hom}%
\left( \mathcal{H}\left( V_{1}\right) ,\mathcal{H}\left( V_{2}\right)
\right) $. The third property implies that every oriented Lagrangian $%
L^{\circ }$ in $\overline{V}_{1}\times V_{2}$ (i.e., oriented canonical
relation from $V_{1}$ to $V_{2}$ (cf. \cite{We81, We82})) can be quantized into a well defined
operator
\begin{eqnarray}
L^{\circ }\longmapsto A_{L^{\circ }}\in \mathrm{Hom}\left( \mathcal{H}\left(
V_{1}\right) ,\mathcal{H}\left( V_{2}\right) \right) . \nonumber
\end{eqnarray}

In this regard, a particular kind of oriented Lagrangian in 
$\overline{V}\times V$ is the graph $\Gamma _{g}$ of a symplectic linear map $%
g:V\rightarrow V$, \ $g\in Sp\left( V\right) $. The orientation is
automatic in this case - it is induced from $\omega ^{\wedge -n}$, $\textrm{dim} (V)=2n$%
, through the isomorphism $p_{V}:\Gamma _{g}{\rightarrow }V$%
, where $p_{V}:\overline{V}\times V\rightarrow V$ is the projection on the $%
V$-coordinate.

A further and more detailed study of these properties will appear in a
subsequent work.

\subsubsection{The strong Stone-von Neumann theorem}

The main technical result of these notes is a proof (\cite{GH05, H06}
unpublished) of a stronger form of the Stone-von Neumann theorem for the
Heisenberg group over $\mathbb{F}_{q}$. In this regard we describe an
algebro-geometric object (an $\ell $-adic perverse Weil sheaf $\mathcal{K}$%
), which, in particular, implies the strong Stone-von Neumann theorem. The
construction of the sheaf $\mathcal{K}$ is one of the main contributions of
this work.

Finally, we note that our result answers, for the case of the Heisenberg
group, a question of Kazhdan \cite{K04} about the possible existence of a
canonical Hilbert space attached to a coadjoint orbit of a general unipotent
group over $\mathbb{F}_{q}$.

We devote the rest of the introduction to an intuitive explanation of the
main ideas and results of these notes.

\subsection{Quantization of symplectic vectors spaces over finite fields}

Let $\left( V,\omega \right) $ be a $2n$-dimensional symplectic vector space
over the finite field $\mathbb{F}_{q}$, assuming $q$ is odd. The vector
space $V$ considered as an abelian group admits a non-trivial central
extension $H$, called the \textit{Heisenberg group}, which can be presented
as $H=V\times \mathbb{F}_{q}$ with center $Z=Z(H)=\left \{ \left( 0,z\right)
:z\in \mathbb{F}_{q}\right \} $. The group $Sp=Sp\left( V\right) $ acts on $%
H $ by group automorphisms via its tautological action on the $V$-coordinate.

The celebrated Stone-von Neumann theorem \cite{N31, St30} asserts that given
a non-trivial central character $\psi :Z\rightarrow
%TCIMACRO{\U{2102} }%
%BeginExpansion
\mathbb{C}
%EndExpansion
^{\times }$, there exists a unique (up to isomorphism) irreducible
representation $\pi :H\rightarrow GL(\mathcal{H})$ such that the center acts
by $\psi $, i.e., $\pi _{|Z}=\psi \cdot \textrm{Id}_{\mathcal{H}}$. The
representation $\pi $ is called the \textit{Heisenberg representation. }

Choosing a Lagrangian\textit{\ }subvector space $L\in
\textrm{Lag}\left( V\right) $ (the set $\textrm{Lag}\left( V\right)$ is called the \textit{Lagrangian Grassmanian}) we can define a model $\left( \pi _{L},H,\mathcal{H}%
_{L}\right) $ of the Heisenberg representation, where $\mathcal{H}_{L}$
consists of functions $f:H\rightarrow
%TCIMACRO{\U{2102} }%
%BeginExpansion
\mathbb{C}
%EndExpansion
$ satisfying $f(z\cdot l\cdot h)=\psi (z)f(h)$ for every $l\in L$, $\ z\in Z$
and the action $\pi _{L}$ is given by right translation. The problematic
issue in this construction is that there is no preferred choice of a
Lagrangian subspace $L\in V$ and consequently none of the spaces $\mathcal{H}%
_{L}$ admit an action of the group $Sp$. In fact, an element $g\in Sp$
induces an isomorphism
\begin{equation}
g:\mathcal{H}_{L}{\rightarrow }\mathcal{H}_{gL},
\label{equiv_eq}
\end{equation}
for every $L\in \textrm{Lag}\left( V\right) $.

\subsubsection{The strong Stone-von Neumann theorem}

The strategy that we will employ is: \textquotedblleft If you can not choose
a preferred Lagrangian subspace then work with all of them
simultaneously\textquotedblright .

We can think of the system of models $\left \{ \mathcal{H}_{L}\right \} $ as
a vector bundle $\mathfrak{H}$ on $\textrm{Lag}$ with fibers $\mathfrak{H}_{|L}=%
\mathcal{H}_{L}$, the condition (\ref{equiv_eq}) means that $\mathfrak{H}$
is equipped with an $Sp$-equivariant structure and what we seek is a
canonical trivialization of $\mathfrak{H}$. More formally, we seek for a
canonical system\textit{\ }of intertwining morphisms\textit{\ }$F_{M,L}\in
\mathrm{Hom}_{H}\left( \mathcal{H}_{L},\mathcal{H}_{M}\right) $, for every $%
L,M\in \textrm{Lag}\left( V\right) $. The existence of such a system is the content
of the strong Stone-von Neumann theorem.

\begin{theorem}[Strong Stone-von Neumann theorem]
There exists a canonical system of intertwining morphisms $\{F_{M,L}\in
\mathrm{Hom}_{H}\left( \mathcal{H}_{L},\mathcal{H}_{M}\right) \}$ satisfying
the multiplicativity property $F_{N,M}\circ F_{M,L}=F_{N,L}$, for every $%
N,M,L\in \textrm{Lag}\left( V\right) $.
\end{theorem}

\begin{remark}[Important remark]
It is important to note here that the precise statement involves the finer
notion of an oriented Lagrangian subspace \cite{B04, LV80}, but for the sake
of the introduction we will ignore this technical nuance.
\end{remark}

The Hilbert space $\mathcal{H}\left( V\right) $ consists of \ systems of
vectors $\left( v_{L}\in \mathcal{H}_{L}\right) _{L\in \textrm{Lag}}$ such that $%
F_{M,L}\left( v_{L}\right) =v_{M}$, for every $L,M\in \textrm{Lag}\left( V\right) $.
The vector space $\mathcal{H}\left( V\right) $ can be thought of as the
space of horizontal sections\ of $\mathfrak{H}$.

As it turns out, the symplectic group $Sp$ naturally acts on $\mathcal{H}%
\left( V\right) $. We denote this representation by $\left( \rho _{V},Sp,%
\mathcal{H}\left( V\right) \right) $, and refer to it as the \textit{%
canonical model }of the Weil representation. We proceed to explain the main
underlying idea behind the construction of the system $\{F_{M,L}\}$.

\subsection{Canonical system of intertwining morphisms}

The construction will be close in spirit to the procedure of
\textquotedblleft analytic continuation\textquotedblright. We consider the
subset $U\subset \textrm{Lag}\left( V\right) ^{2}$, consisting of pairs $(L,M)\in
\textrm{Lag}\left( V\right)^{2}$ which are in general position, that is $L\cap M=0$.
The basic idea is that for a pair $\left( L,M\right) \in U$, $F_{M,L}$ can
be given by an explicit formula - \textbf{ansatz}. The main statement is
that this formula admits a unique multiplicative extension to the set of all
pairs. The extension is constructed using algebraic geometry.

\subsubsection{Extension to singular pairs}

It will be convenient to work in the setting of \textit{kernels.} In more
detail, every intertwining morphism $F\in $ $\mathrm{Hom}_{H}\left(
\mathcal{H}_{L},\mathcal{H}_{M}\right) $ can be presented by a kernel
function $K\in
%TCIMACRO{\U{2102} }%
%BeginExpansion
\mathbb{C}
%EndExpansion
\left( H,\psi \right)$ satisfying $K\left( m\cdot h\cdot l\right) =K\left(
h\right) $, for every $m\in M$ and $l\in L$ (we denote by $\mathbb{C}(H,\psi)$ the subspace of functions $f\in
%TCIMACRO{\U{2102} }%
%BeginExpansion
\mathbb{C}
%EndExpansion
\left( H\right) $ which are $\psi $-equivariant with respect to the center,
that is $f\left( z\cdot h\right) =\psi \left( z\right) f\left( h\right) $,
for every $z\in Z$). Moreover, this presentation is
unique when $\left( M,L\right) \in U$; hence, in this case, we have a unique
kernel $K_{M,L}$ representing our given $F_{M,L}$. If we denote by $O$ the
set $U\times H$, we see that the collection $\left \{ K_{M,L}:\left(
M,L\right) \in U\right \} $ forms a function $K_{O}\in
%TCIMACRO{\U{2102} }%
%BeginExpansion
\mathbb{C}
%EndExpansion
\left( O\right) $ given by $K_{O}\left( M,L\right) =K_{M,L}$ for every $\left(
M,L\right) \in U$. The problem is how to (correctly) extend the function $%
K_{O}$ to the set $\ X=\textrm{Lag}\left( V\right) ^{2}\times H$. In order to do
that, we invoke the procedure of geometrization, which we briefly explain
below.

\subsubsection{Geometrization}

A general ideology due to Grothendieck is that any meaningful set-theoretic
object is governed by a more fundamental algebro-geometric one. \textit{\ }%
The procedure by which one translates from the set theoretic setting to
algebraic geometry is called \textit{geometrization}, which is a formal
procedure by which sets are replaced by algebraic varieties and functions
are replaced by certain sheaf-theoretic objects.

The precise setting consists of a set $X=\mathbf{X}\left( \mathbb{F}%
_{q}\right) $ of rational points of an algebraic variety $\mathbf{X}$,
defined over $\mathbb{F}_{q}$ and a complex valued function $f\in
%TCIMACRO{\U{2102} }%
%BeginExpansion
\mathbb{C}
%EndExpansion
\left( X\right) $ governed by an $\ell $-adic Weil sheaf $\mathcal{F}$.

The variety $\mathbf{X}$ is a space equipped with an automorphism $Fr:%
\mathbf{X}\rightarrow \mathbf{X}$ \ (called Frobenius), such that the set $X$
is naturally identified with the set of fixed points $X=\mathbf{X}^{Fr}$.

The sheaf $\mathcal{F}$ can be considered as a vector bundle on the variety $%
\mathbf{X}$, equipped with an endomorphism $\theta :\mathcal{F\rightarrow F}$
which lifts $Fr$.

The procedure by which $f$ is obtained from $\mathcal{F}$ is called
Grothendieck's \textit{sheaf-to-function correspondence }and it can be
described, roughly, as follows. Given a point $x\in X$, the endomorphism $%
\theta $ restricts to an endomorphism $\theta _{x}:\mathcal{F}%
_{|x}\rightarrow \mathcal{F}_{|x}$ of the fiber $\mathcal{F}_{|x}$. The
value of $f$ on the point $x$ is defined to be
\begin{eqnarray}
f(x)=\textrm{Tr}(\theta _{x}:\mathcal{F}_{|x}\rightarrow \mathcal{F}_{|x}). \nonumber
\end{eqnarray}%
The function defined by this procedure is denoted by $f=f^{\mathcal{F}}$.

\subsubsection{Solution to the extension problem}

Our extension problem fits nicely into the geometrization setting: The sets $%
O,X$ are sets of rational points of corresponding algebraic varieties $%
\mathbf{O,X}$, the imbedding $j:O\hookrightarrow X$ is induced from an open
imbedding $j:\mathbf{O\hookrightarrow X}$ and, finally, the function $K_{O}$
comes from a Weil sheaf $\mathcal{K}_{\mathbf{O}}$ on the variety $\mathbf{O}
$.

The extension problem is solved as follows: First extend the sheaf $\mathcal{%
K}_{\mathbf{O}}$ to a sheaf $\mathcal{K}$ on the variety $\mathbf{X}$ and
then take the corresponding function $K=f^{\mathcal{K}}$, which establishes
the desired extension. The reasoning behind this strategy is that in the
realm of sheaves there exist several functorial operations of extension,
probably the most interesting one is called \textit{perverse extension} {\cite{BBD82}. 
The sheaf $\mathcal{K}$ is defined as the perverse
extension of $\mathcal{K}_{\mathbf{O}}$.

\subsection{Structure of the notes}

Apart from the introduction, the notes consists of three sections.

In Section \ref{pre_sec}, all basic constructions are introduced and main
statements are formulated. We begin with the definition of the Heisenberg
group and the Heisenberg representation. Next, we introduce the
canonical system of intertwining morphisms between different models of the
Heisenberg representation and formulate the strong Stone von-Neumann theorem
(Theorem \ref{SSv-N_thm}). We proceed to explain how to present an
intertwining morphism by a kernel function, and we reformulate the strong
Stone von-Neumann theorem in the setting of kernels (Theorem \ref{SSv-N2_thm}%
). Using Theorem \ref{SSv-N_thm}, we construct a quantization functor $%
\mathcal{H}$. We finish this section by showing that $\mathcal{H}$ is a
monoidal functor and that it is compatible with duality and the operation of
linear symplectic reduction. In section \ref{geom_sec}, we construct a sheaf
theoretic counterpart for the canonical system of intertwining morphisms
(Theorem \ref{GIS_thm}). This sheaf is then used to prove Theorem \ref%
{SSv-N2_thm}. Finally, in Section \ref{GIS_sec} we sketch the proof of
Theorem \ref{GIS_thm}. Complete proofs for the statements appearing in these
notes will appear elsewhere.

\subsection{Acknowledgments}

We would like to thank our scientific advisor J. Bernstein for his interest
and guidance, and for his idea about the notion of oriented Lagrangian
subspace. It is a pleasure to thank D. Kazhdan for sharing with us his
thoughts about the possible existence of canonical Hilbert spaces. We thank
A. Weinstein for teaching us some of his ideas concerning quantization, and
for the opportunity to present this work in the symplectic geometry seminar,
Berkeley, February 2007. We acknowledge O. Gabber for his remark concerning
the characterization of the Weil representation. We would like to
acknowledge M. Vergne for her encouragement. Finally, we would like to thank
O. Ceyhan and the organizers of the conference AGAQ, held in Istanbul during
June 2006, for the invitation to present this work.

\section{Quantization of symplectic vector spaces over finite fields\label%
{pre_sec}}

\subsection{The Heisenberg group}

Let $(V,\omega )$ be a $2n$-dimensional symplectic vector space over the
finite field $\mathbb{F}_{q}$. Considering $V$ as an abelian group, it
admits a non-trivial central extension called the \textit{Heisenberg }group.
\ Concretely, the group $H=H\left( V\right) $ can be presented as the set $%
H=V\times \mathbb{F}_{q}$ with the multiplication given by%
\begin{eqnarray}
(v,z)\cdot (v^{\prime },z^{\prime })=(v+v^{\prime },z+z^{\prime }+\frac{1}{2}\omega (v,v^{\prime })). \nonumber
\end{eqnarray}

The center of $H$ is $\ Z=Z(H)=\left \{ (0,z):z\in \mathbb{F}%
_{q}\right \} .$ The symplectic group $Sp=Sp(V)$ acts by automorphism of $H$
through its tautological action on the $V$-coordinate.

\subsection{The Heisenberg representation \label{HR}}

One of the most important attributes of the group $H$ is that it admits,
principally, a unique irreducible representation. We will call this property
\textit{The Stone-von Neumann property} (S-vN for short). The precise statement
goes as follows. Let $\psi :Z\rightarrow
%TCIMACRO{\U{2102} }%
%BeginExpansion
\mathbb{C}
%EndExpansion
^{\times }$ be a non-trivial character of the center. For example we can
take $\psi \left( z\right) =e^{\frac{2\pi i}{p}\textrm{tr}\left( z\right) }$. It is
not hard to show

\begin{theorem}[Stone-von Neumann property]
\label{S-vN_thm}There exists a unique (up to isomorphism) irreducible
unitary representation $(\pi ,H,\mathcal{H)}$ with the center acting by $%
\psi ,$ i.e., $\pi _{|Z}=\psi \cdot \textrm{Id}_{\mathcal{H}}$.
\end{theorem}

The representation $\pi $ which appears in the above theorem will be called
the \textit{Heisenberg representation}.

\subsection{The strong Stone-von Neumann property}

Although the representation $\pi$ is unique, it admits a multitude of
different models (realizations); in fact this is one of its most interesting
and powerful attributes. These models appear in families. In this work we
will be interested in a particular family of such models which are
associated with Lagrangian subspaces in $V$.

Let us denote by $\textrm{Lag}=\textrm{Lag}\left( V\right) $ the set of Lagrangian subspaces
in $V$. Let $%
%TCIMACRO{\U{2102} }%
%BeginExpansion
\mathbb{C}
%EndExpansion
\left( H,\psi \right) $ denote the subspace of functions $f$ $\in
%TCIMACRO{\U{2102} }%
%BeginExpansion
\mathbb{C}
%EndExpansion
\left( H\right) $, satisfying the equivariance property $f\left( z\cdot
h\right) =\psi \left( z\right) f\left( h\right) $, for every $z\in Z$.

Given a Lagrangian subspace $L\in \textrm{Lag}$, we can construct a model $(\pi
_{L},H,\mathcal{H}_{L})$ of the Heisenberg representation: The vector space $%
\mathcal{H}_{L}$ consists of functions $f\in
%TCIMACRO{\U{2102} }%
%BeginExpansion
\mathbb{C}
%EndExpansion
\left( H,\psi \right) $ satisfying $f\left( l\cdot h\right) =f\left(
h\right) $, for every $l\in L$ and the Heisenberg action is given by right
translation $\left( \pi_{L}\left( h\right) \triangleright f\right)
\left( h^{\prime }\right) =f\left( h^{\prime }\cdot h\right) $, for $f\in
\mathcal{H}_{L}$.

\begin{definition}
An \underline{oriented Lagrangian} $L^{\circ }$ is a pair $L^{\circ }=\left(
L,o_{L}\right) $, where $L$ is a Lagrangian subspace in $V$ and $o_{L}$ is a
non-zero vector in $\bigwedge \nolimits^{\textrm{top}}L$.
\end{definition}

Let $\textrm{Lag}^{\circ }=\textrm{Lag}^{\circ }\left( V\right) $ denote the set of oriented
Lagrangian subspaces in $V$. We associate to each oriented Lagrangian
subspace $L^{\circ }$, a model $(\pi _{L^{\circ }},H,\mathcal{H}_{L^{\circ
}})$ of the Heisenberg representation simply by forgetting the orientation, taking $\mathcal{H}_{L^{\circ }}=\mathcal{H}_{L}$ and $\pi _{L^{\circ
}}=\pi _{L}$. Sometimes, we will use a more informative notation $\mathcal{H}%
_{L^{\circ }}=\mathcal{H}_{L^{\circ }}\left( V\right) $ or $\mathcal{H}%
_{L^{\circ }}=\mathcal{H}_{L^{\circ }}\left( V,\psi \right) $.

\subsubsection{Canonical system of intertwining morphisms \label{CIM_subsub}}

Given a pair $\left( M^{\circ },L^{\circ }\right) \in \textrm{Lag}^{\circ 2}$, the
models $\mathcal{H}_{L^{\circ }}$ and $\mathcal{H}_{M^{\circ }}$ are
isomorphic as representations of $H$ by Theorem \ref{S-vN_thm}, moreover,
since the Heisenberg representation is irreducible, the vector space $%
\mathrm{Hom}_{H}\left( \mathcal{H}_{L^{\circ }},\mathcal{H}_{M}^{\circ
}\right) $ of intertwining morphisms is one-dimensional. Roughly, the strong
Stone-von Neumann property asserts the existence of a distinguished element $%
F_{M^{\circ },L^{\circ }}\in $ $\mathrm{Hom}_{H}\left( \mathcal{H}_{L^{\circ
}},\mathcal{H}_{M}^{\circ }\right) $, for every pair $\left( M^{\circ
},L^{\circ }\right) \in \textrm{Lag}^{\circ 2}$. The precise statement involves the
following definition:

\begin{definition}
A system $\{F_{M^{\circ },L^{\circ }}\in \mathrm{Hom}_{H}\left( \mathcal{H}%
_{L^{\circ }},\mathcal{H}_{M}^{\circ }\right) :\left( M^{\circ },L^{\circ
}\right) \in \textrm{Lag}^{\circ 2}\}$ of intertwining morphisms is called \underline{%
multiplicative} if for every triple $\left( N^{\circ },M^{\circ },L^{\circ
}\right) \in \textrm{Lag}^{\circ 3}$ the following equation holds
\begin{eqnarray}
F_{N^{\circ },L^{\circ }}=F_{N^{\circ },M^{\circ }}\circ F_{M^{\circ
},L^{\circ }}. \nonumber
\end{eqnarray}
\end{definition}

We proceed as follows. Let $U\subset \textrm{Lag}^{\circ 2}$ denote the set of pairs $%
\left( M^{\circ },L^{\circ }\right) \in \textrm{Lag}^{\circ 2}$ which are in general
position, i.e., $L\cap M=0$. For $\left( M^{\circ },L^{\circ }\right) \in
U $, we define $F_{M^{\circ },L^{\circ }}$ by the following explicit formula:
\begin{equation}
F_{M^{\circ },L^{\circ }}=C_{M^{\circ },L^{\circ }}\cdot \widetilde{F}_{M,L},
\label{ansatz}
\end{equation}%
where $\widetilde{F}_{M,L}:\mathcal{H}_{L^{\circ }}\rightarrow \mathcal{H}%
_{M^{\circ }}$ is the averaging morphism%
\begin{eqnarray}
\widetilde{F}_{M,L}\left[ f\right] \left( h\right) =\sum \limits_{m\in
M}f\left( m\cdot h\right) , \nonumber
\end{eqnarray}%
for every $f\in \mathcal{H}_{L^{\circ }}$ and $C_{M^{\circ },L^{\circ }}$ is
a normalization constant given by
\begin{eqnarray}
C_{M^{\circ },L^{\circ }}=(G_{1}/q)^n\cdot\sigma \left((-1)^{n\choose 2}\omega _{\wedge }\left( o_{L},o_{M}\right) \right),
\nonumber
\end{eqnarray}
where $n=\frac{\textrm{dim}(V)}{2}$, $\sigma $ is the unique quadratic character (also called the Legendre
character) of the multiplicative group $G_{m}=\mathbb{F}_{q}^{\times }$, $%
G_{1}$ is the one-dimensional Gauss sum
\begin{eqnarray}
G_{1}=\sum \limits_{z\in \mathbb{F}_{q}}\psi \left( \frac{1}{2}z^{2}\right)
, \nonumber
\end{eqnarray}%
and $\omega _{\wedge }$ is the pairing $\omega _{\wedge
}:\bigwedge^{\textrm{top}}L\bigotimes $ $\bigwedge^{\textrm{top}}M\rightarrow $ $\mathbb{F}%
_{q} $ induced by the symplectic form.

\begin{theorem}[The strong Stone - von Neumann property]
\label{SSv-N_thm}There exists a unique system $\{F_{M^{\circ },L^{\circ }}\}$
of intertwining morphisms satisfying

\begin{enumerate}
\item Restriction. \label{res_prop}For every pair $\left( M^{\circ
},L^{\circ }\right) \in U$, $F_{M^{\circ },L^{\circ }}$ is given by (\ref%
{ansatz}).

\item Multiplicativity. \label{mult_prop}For every triple $\left( N^{\circ
},M^{\circ },L^{\circ }\right) \in \textrm{Lag}^{\circ 3}$,
\begin{eqnarray}
F_{N^{\circ },L^{\circ }}=F_{N^{\circ },M^{\circ }}\circ F_{M^{\circ
},L^{\circ }}. \nonumber
\end{eqnarray}
\end{enumerate}
\end{theorem}

Theorem \ref{SSv-N_thm} will follow from Theorem \ref{SSv-N2_thm} below.

Granting the existence and uniqueness of the system $\left \{ F_{M^{\circ
},L^{\circ }}\right \} $, we can write $F_{M^{\circ },L^{\circ }}$\ in a
closed form, for a general pair $\left( M^{\circ },L^{\circ }\right) \in
\textrm{Lag}^{\circ 2}$. In order to do that we need to fix some additional
terminology.

Let $I=M\cap L$. We have canonical tensor product decompositions
\begin{eqnarray*}
\bigwedge \nolimits^{\textrm{top}}M &=&\bigwedge \nolimits^{\textrm{top}}I\bigotimes
\bigwedge \nolimits^{\textrm{top}}M/I, \\
\bigwedge \nolimits^{\textrm{top}}L &=&\bigwedge \nolimits^{\textrm{top}}I\bigotimes
\bigwedge \nolimits^{\textrm{top}}L/I.
\end{eqnarray*}

In terms of the above decompositions, the orientation can be written
in the form $o_{M}=\iota _{M}\otimes o_{M/I}$, $o_{L}=\iota _{L}\otimes
o_{L/I}$. Using the same notations as before,\ we denote by $\widetilde{F}%
_{M,L}:\mathcal{H}_{L^{\circ }}\rightarrow \mathcal{H}_{M^{\circ }}$ the
averaging morphism
\begin{eqnarray}
\widetilde{F}_{M,L}\left[ f\right] \left( h\right) =\sum \limits_{\overline{m%
}\in M/I}f\left( m\cdot h\right) , \nonumber
\end{eqnarray}%
for $f\in \mathcal{H}_{L^{\circ }}$ and by $C_{M^{\circ },L^{\circ }}$ the
normalization constant
\begin{eqnarray}
C_{M^{\circ },L^{\circ }}=(G_{1})^k \cdot \sigma \left((-1)^{k \choose 2} \frac{\iota _{M}}{\iota _{L}}\cdot \omega _{\wedge
}\left( o_{L/I},o_{M/I}\right) \right), \nonumber
\end{eqnarray}
where $k=\frac{\textrm{dim}(I^\perp/I)}{2}$.

\begin{proposition}
\label{explicit_prop} For every $\left( M^{\circ },L^{\circ }\right) \in
\textrm{Lag}^{\circ 2}$
\begin{eqnarray}
F_{M^{\circ },L^{\circ }}=C_{M^{\circ },L^{\circ }}\cdot \widetilde{F}_{M,L}.
 \nonumber
\end{eqnarray}
\end{proposition}

\subsection{Kernel presentation of an intertwining morphism}

An explicit way to present an intertwining morphism is via a kernel
function. Fix a pair $\left( M^{\circ },L^{\circ }\right) \in \textrm{Lag}^{\circ 2}$
and let $%
%TCIMACRO{\U{2102} }%
%BeginExpansion
\mathbb{C}
%EndExpansion
\left( M\backslash H/L,\psi \right) $ denote the subspace of functions $f$ $%
\in
%TCIMACRO{\U{2102} }%
%BeginExpansion
\mathbb{C}
%EndExpansion
\left( H,\psi \right) $ satisfying the equivariance property $f\left( m\cdot
h\cdot l\right) =f\left( h\right) $ for every $m\in M$ and $l\in L$. Given a
function $K\in
%TCIMACRO{\U{2102} }%
%BeginExpansion
\mathbb{C}
%EndExpansion
\left( M\backslash H/L,\psi \right) $, we can associate to it an
intertwining morphism $I\left[ K\right] \in $ $\mathrm{Hom}_{H}(\mathcal{H}%
_{L^{\circ }},\mathcal{H}_{M}^{\circ })$ defined by
\begin{eqnarray}
I\left[ K\right] \left( f\right) =K\ast f=m_{!}\left( K\boxtimes _{Z\cdot
M}f\right) , \nonumber
\end{eqnarray}%
for\ every $f\in \mathcal{H}_{L^{\circ }}$. Here, $K\boxtimes _{Z\cdot L}f$ $%
\ $denotes the function $K\boxtimes f\in
%TCIMACRO{\U{2102} }%
%BeginExpansion
\mathbb{C}
%EndExpansion
\left( H\times H\right) $, factored to the quotient $H\times _{Z\cdot L}H$
and $m_{!}$ denotes the operation of summation along the fibers of the
multiplication mapping $m:H\times H\rightarrow H$. The function $K$ is
called an\textit{\ intertwining kernel}. The procedure just described
defines a linear transform
\begin{eqnarray}
I:%
%TCIMACRO{\U{2102} }%
%BeginExpansion
\mathbb{C}
%EndExpansion
\left( M\backslash H/L,\psi \right) \longrightarrow \mathrm{Hom}_{H}(%
\mathcal{H}_{L^{\circ }},\mathcal{H}_{M^{\circ }}). \nonumber
\end{eqnarray}

An easy verification reveals that $I$ is surjective, but it is injective
only when $M,L$ are in general position.

Fix a triple $\left( N^{\circ },M^{\circ },L^{\circ }\right) \in \textrm{Lag}^{\circ
3}$. Given kernels $K_{1}\in
%TCIMACRO{\U{2102} }%
%BeginExpansion
\mathbb{C}
%EndExpansion
\left( N\backslash H/M,\psi \right) $ and $K_{2}\in
%TCIMACRO{\U{2102} }%
%BeginExpansion
\mathbb{C}
%EndExpansion
\left( M\backslash H/L,\psi \right) $, their convolution $K_{1}\ast
K_{2}=m_{!}\left( K_{1}\boxtimes _{Z\cdot M}K_{2}\right) $ lies in $%
%TCIMACRO{\U{2102} }%
%BeginExpansion
\mathbb{C}
%EndExpansion
\left( N\backslash H/L,\psi \right) $. The transform $I$ sends convolution
of kernels to composition of operators
\begin{eqnarray}
I\left[ K_{1}\ast K_{2}\right] =I\left[ K_{1}\right] \circ I\left[ K_{2}%
\right] . \nonumber
\end{eqnarray}

\subsubsection{Canonical system of intertwining kernels \label%
{sys_kernels_subsub}}

Below, we formulate a slightly stronger version of Theorem \ref{SSv-N_thm},
in the setting of kernels.

\begin{definition}
A system $\{K_{M^{\circ },L^{\circ }}\in
%TCIMACRO{\U{2102} }%
%BeginExpansion
\mathbb{C}
%EndExpansion
\left( M\backslash H/L,\psi \right) :\left( M^{\circ },L^{\circ }\right) \in
\textrm{Lag}^{\circ 2}\}$ of kernels is called \underline{multiplicative} if for
every triple $\left( N^{\circ },M^{\circ },L^{\circ }\right) \in \textrm{Lag}^{\circ
3}$ the following equation holds
\begin{eqnarray}
K_{N^{\circ },L^{\circ }}=K_{N^{\circ },M^{\circ }}\ast K_{M^{\circ
},L^{\circ }} \nonumber
\end{eqnarray}
\end{definition}

A multiplicative system of kernels $\{K_{M^{\circ },L^{\circ }}\}$ can be
equivalently thought of as a single function $K\in
%TCIMACRO{\U{2102} }%
%BeginExpansion
\mathbb{C}
%EndExpansion
\left( \textrm{Lag}^{\circ 2}\times H\right) $, $K\left( M^{\circ },L^{\circ }\right)
=K_{M^{\circ },L^{\circ }}$, satisfying\ the following multiplicativity
relation on $\textrm{Lag}^{\circ 3}\times H$
\begin{equation}
p_{12}^{\ast }K\ast p_{23}^{\ast }K=p_{13}^{\ast }K, \label{mult_eq}
\end{equation}%
where $p_{ij}\left( \left( L_{1}^{\circ },L_{2}^{\circ },L_{3}^{\circ
}\right) ,h\right) =\left( \left( L_{i}^{\circ },L_{j}^{\circ }\right)
,h\right) $ are the projections on the $i,j$ copies of $\textrm{Lag}^{\circ }$ and
the left-hand side of (\ref{mult_eq}) means fiberwise convolution, namely $%
p_{12}^{\ast }K\ast p_{23}^{\ast }K(L_{1}^{\circ },L_{2}^{\circ
},L_{3}^{\circ })=K\left( L_{1}^{\circ },L_{2}^{\circ }\right) \ast K\left(
L_{2}^{\circ },L_{3}^{\circ }\right) $. To simplify notations, we will
sometimes suppress the projections $p_{ij}$ from (\ref{mult_eq}) obtaining
a much cleaner formula%
\begin{eqnarray}
K\ast K=K. \nonumber
\end{eqnarray}

We proceed along lines similar to Section \ref{CIM_subsub}. For every $%
\left( M^{\circ },L^{\circ }\right) \in U$, there exists a unique kernel $%
K_{M^{\circ },L^{\circ }}\in
%TCIMACRO{\U{2102} }%
%BeginExpansion
\mathbb{C}
%EndExpansion
\left( M\backslash H/L,\psi \right) $ such that $F_{M^{\circ },L^{\circ }}=I%
\left[ K_{M^{\circ },L^{\circ }}\right] $, which is given by the following
explicit formula%
\begin{equation}
K_{M^{\circ },L^{\circ }}=C_{M^{\circ },L^{\circ }}\cdot \widetilde{K}%
_{M^{\circ },L^{\circ }},  \label{ansatz2}
\end{equation}%
where $\widetilde{K}_{M^{\circ },L^{\circ }}=\left( \iota ^{-1}\right)
^{\ast }\psi $, $\iota =\iota _{M^{\circ },L^{\circ }}$ is the isomorphism
given by the composition $Z\hookrightarrow H\twoheadrightarrow M\backslash
H/L$. The system $\{K_{M^{\circ },L^{\circ }}:\left( M^{\circ },L^{\circ
}\right) \in U\}$ yields a well defined function $K_{U}\in
%TCIMACRO{\U{2102} }%
%BeginExpansion
\mathbb{C}
%EndExpansion
\left( U\times H\right) $.

\begin{theorem}[Canonical system of kernels]
\label{SSv-N2_thm}There exists a unique function $K\in
%TCIMACRO{\U{2102} }%
%BeginExpansion
\mathbb{C}
%EndExpansion
\left( \textrm{Lag}^{\circ 2}\times H\right) $ satisfying

\begin{enumerate}
\item Restriction. $K_{|U}=K_{U}$.

\item Multiplicativity. \label{mult_prop2}$K\ast K=K$.
\end{enumerate}
\end{theorem}

We note that the proof of the uniqueness part in Theorem \ref{SSv-N2_thm} is
easy, it follows from the fact that for every pair $N^{\circ },L^{\circ }\in
\textrm{Lag}^{\circ }$ one can find a third $M^{\circ }\in \textrm{Lag}^{\circ }$ such that
the pairs $N^{\circ },M^{\circ }$ and $M^{\circ },L^{\circ }$ are in general
position. Therefore, by the multiplicativity property (Property \ref%
{mult_prop2}), $K_{N^{\circ },L^{\circ }}=K_{N^{\circ },M^{\circ }}\ast
K_{M^{\circ },L^{\circ }}$. The proof of the existence part will be
algebro-geometric (see Section \ref{geom_sec}). Finally, we note that
Theorem \ref{SSv-N_thm} follows from Theorem \ref{SSv-N2_thm} by taking $%
F_{M^{\circ },L^{\circ }}=I\left[ K_{M^{\circ },L^{\circ }}\right] $.

\subsection{The canonical vector space\label{CH_subsec}}

Let us denote by $\mathsf{Symp}$ the category whose objects are symplectic
vector spaces over $\mathbb{F}_{q}$ and morphisms are linear isomorphisms of
symplectic vector spaces. Using the canonical system of intertwining
morphisms $\left \{ F_{M^{\circ },L^{\circ }}\right \} $ we can associate,
in a functorial manner, a vector space $\mathcal{H}\left( V\right) $ \ to a
symplectic vector space $V\in \mathsf{Symp}$. The construction proceeds as
follows.

Let $\Gamma \left( V\right) $ denote the total vector space%
\begin{eqnarray}
\Gamma (V)=\bigoplus \limits_{L^{\circ }\in \textrm{Lag}^{\circ }\left( V\right) }%
\mathcal{H}_{L^{\circ }}, \nonumber
\end{eqnarray}

Define $\mathcal{H}\left( V\right) $ to be the subvector space of $\  \Gamma
\left( V\right) $ consisting of sequences $(v_{L^{\circ }}\in \mathcal{H}%
_{L^{\circ }}:L^{\circ }\in \textrm{Lag}^{\circ })$ satisfying $F_{M^{\circ
},L^{\circ }}\left( v_{L^{\circ }}\right) =v_{M^{\circ }}$ for every $\left(
M^{\circ },L^{\circ }\right) \in \textrm{Lag}^{\circ 2}\left( V\right) $. We will
call the vector space $\mathcal{H}\left( V\right) $ the \textit{canonical
vector space} associated with $V$. Sometimes we will use the more
informative notation $\mathcal{H}\left( V\right) =\mathcal{H}\left( V,\psi
\right)$.

\begin{proposition}[Functoriality]
\label{functor_prop}The rule $V\mapsto \mathcal{H}\left( V\right) $
establishes a contravariant (quantization) functor
\begin{eqnarray}
\mathcal{H}:\mathsf{Symp}\longrightarrow \mathsf{Vect,} \nonumber
\end{eqnarray}%
where $\mathsf{Vect}$ denote the category of finite dimensional complex
vector spaces.
\end{proposition}

Considering a fixed symplectic vector space $V$, we obtain as a consequence
a representation $\left( \rho _{V},Sp\left( V\right) ,\mathcal{H}\left(
V\right) \right) $, with $\rho _{V}\left( g\right) =\mathcal{H}\left(
g^{-1}\right) $, for every $g\in Sp\left( V\right) $. The representation $%
\rho _{V}$ is isomorphic to the Weil representation and we call it the
\textit{canonical model} of the Weil representation.

\begin{remark}
The canonical model $\rho _{V}$ can be viewed from another perspective: \ We
begin with the total vector space $\Gamma $ and make the following two
observations. First observation is that the symplectic group $Sp$ acts
naturally on $\Gamma $, the action is of a geometric nature, i.e., induced from
the diagonal action on $\textrm{Lag}^{\circ }\times H$. Second observation is that
the system $\left \{ F_{M^{\circ },L^{\circ }}\right \} $ defines an $Sp$%
-invariant idempotent (total Fourier transform) $F:\Gamma \rightarrow \Gamma
$ given by
\begin{eqnarray}
F\left( v_{L^{\circ }}\right) =\frac{1}{\# \left( \textrm{Lag}^{\circ }\right) }%
\bigoplus \limits_{M^{\circ }\in \textrm{Lag}^{\circ }}F_{M^{\circ },L^{\circ
}}\left( v_{L^{\circ }}\right) , \nonumber
\end{eqnarray}%
for every $L^{\circ }\in \textrm{Lag}^{\circ }$ and $v_{L^{\circ }}\in \mathcal{H}%
_{L^{\circ }}$. The situation is summarized in the following diagram:
\begin{eqnarray}
Sp\circlearrowright \Gamma \circlearrowleft F. \nonumber
\end{eqnarray}%
The canonical model is given by the image of $F$, that is, $\mathcal{H}%
\left( V\right) =F\Gamma $. The nice thing about this point of view is that
it shows a clear distinction between operators associated with action of the
symplectic group and operators associated with intertwining morphisms.
Finally, we remark that one can also consider the $Sp$-invariant
idempotent $F^{\bot }=\textrm{Id}-F$ and the associated representation $\left( \rho
_{V}^{\bot },Sp,\mathcal{H}\left( V\right) ^{\bot }\right) $, with $\mathcal{%
H}\left( V\right) ^{\bot }=F^{\bot }\Gamma $. The meaning of this
representation is unclear.
\end{remark}

\subsubsection{Compatibility with Cartesian products}

The category $\mathsf{Symp}$ \ admits a monoidal structure given by
Cartesian product of symplectic vector spaces. The category $\mathsf{Vect}$
admits the standard monoidal structure given by tensor product. With respect
to these monoidal structures, the functor $\mathcal{H}$ is a monoidal
functor.

\begin{proposition}
\label{Cartesian_prop}There exists a natural isomorphism
\begin{eqnarray}
\alpha _{V_{1}\times V_{2}}:\mathcal{H}\left( V_{1}\times V_{2}\right)
{\rightarrow }\mathcal{H}\left( V_{1}\right) \otimes
\mathcal{H}\left( V_{2}\right) , \nonumber
\end{eqnarray}
where $V_{1},V_{2}\in \mathsf{Symp}$.
\end{proposition}

As a result, we obtain the following compatibility condition between the
canonical models of the Weil representation
\begin{equation}
\alpha _{V_{1}\times V_{2}}:\left( \rho _{V_{1}\times V_{2}}\right)
_{|Sp\left( V_{1}\right) \times Sp\left( V_{2}\right) }{%
\longrightarrow }\rho _{V_{1}}\otimes \rho _{V_{2}}.  \label{product_eq}
\end{equation}

\begin{remark}[\protect \cite{G06}]
Condition (\ref{product_eq}) has an interesting consequence in case the
ground field is $\mathbb{F}_{3}$. In this case, the group $Sp\left( V\right)
$ is not perfect when $\textrm{dim} (V)=2$, therefore, a priori, the Weil
representation is not uniquely defined in this particular situation.
However, since the group $Sp\left( V\right) $ becomes perfect when $\textrm{dim} (V)>2$%
, the canonical model gives a natural choice for the Weil representation in
the \textit{singular} dimension, $\textrm{dim} (V)=2$.
\end{remark}

\subsubsection{Compatibility with symplectic duality}

Let $V=\left( V,\omega \right) \in \mathsf{Symp}$ and let us denote by $%
\overline{V}=\left( V,-\omega \right) $ the symplectic dual of $V$.

\begin{proposition}
\label{duality_prop}There exists a natural non-degenerate pairing
\begin{eqnarray}
\left \langle \cdot ,\cdot \right \rangle _{V}:\mathcal{H}\left( \overline{V}%
,\psi \right) \times \mathcal{H}\left( V,\psi \right) \rightarrow
%TCIMACRO{\U{2102} }%
%BeginExpansion
\mathbb{C}
%EndExpansion
, \nonumber
\end{eqnarray}
where $V\in \mathsf{Symp}$.
\end{proposition}

\subsubsection{Compatibility with symplectic reduction}

Let $V\in \mathsf{Symp}$ and let $I$ be an isotropic subspace in $V$
considered as an abelian subgroup in $H\left( V\right) $. On the one hand,
we can associate to $I$ the subspace $\mathcal{H}\left( V\right) ^{I}$ of $I$%
-invariant vectors. On the other hand, we can form the symplectic reduction $I^{\bot }/I$ and consider the vector space $\mathcal{H}\left( I^{\bot
}/I\right)$ (note that since $I$ is isotropic then $I\subset I^{\perp }$ and $I^{\perp }/I$
is equipped with a natural symplectic structure). 
Roughly, we claim that the vector spaces $\mathcal{H}\left(
I^{\bot }/I\right) $ and $\mathcal{H}\left( V\right) ^{I}$ are naturally
isomorphic. The precise statement involves the following definition

\begin{definition}
An \underline{oriented isotropic subspace} in $V$ is a pair $I^{\circ
}=\left( I,o_{I}\right) $, where $I\subset V$ is an isotropic subspace and $%
o_{I}$ is a non-trivial vector in $\bigwedge^{\textrm{top}}I$.
\end{definition}

\begin{proposition}
\label{reduction_prop} There exists a natural isomorphism%
\begin{eqnarray}
\alpha _{\left( I^{\circ },V\right) }:\mathcal{H}\left( V\right) ^{I}
{\rightarrow }\mathcal{H}\left( I^{\perp }/I\right) , \nonumber
\end{eqnarray}
where, $V\in \mathsf{Symp}$ and $I^{\circ }$ an oriented isotropic subspace
in $V$. The naturality condition is $\mathcal{H}\left( f_{I}\right) \circ
\alpha _{\left( J^{\circ },U\right) }=\alpha _{\left( I^{\circ },V\right)
}\circ \mathcal{H}\left( f\right) ,$ for every $f\in \mathrm{Mor}_{\mathsf{%
Symp}}\left( V,U\right) $ such that $f\left( I^{\circ }\right) =J^{\circ }$
and $f_{I}\in \mathrm{Mor}_{\mathsf{Symp}}\left( I^{\perp }/I,J^{\perp
}/J\right) $ is the induced morphism.
\end{proposition}

As a result we obtain another compatibility condition between the canonical
models of the Weil representation. In order to see this, fix $V\in $ $%
\mathsf{Symp}$ and let $I^{\circ }$ be an oriented isotropic subspace in $V$%
. Let $P\subset Sp\left( V\right) $ be the subgroup of elements $g\in
Sp\left( V\right) $ such that $g\left( I^{\circ }\right) =I^{\circ }$. The
isomorphism $\alpha _{\left( I^{\circ },V\right) }$ establishes the
following isomorphism:
\begin{equation}
\alpha _{\left( I^{\circ },V\right) }:\left( \rho _{V}\right) _{|P}{\longrightarrow }\rho _{I^{\perp }/I}\circ \pi ,
\label{reduction_eq}
\end{equation}%
where $\pi :P\rightarrow Sp\left( I^{\perp }/I\right) $ is the canonical
homomorphism.

\section{Geometric intertwining morphisms\label{geom_sec}}

In this section we are going to prove Theorem \ref{SSv-N2_thm}, by
constructing a geometric counterpart to the set-theoretic system of
intertwining kernels. This will be achieved using geometrization.

\subsection{Preliminaries from algebraic geometry}

We denote by $k$ an algebraic closure of $\mathbb{F}_{q}$. Next we have to take some space to recall notions and notations from algebraic geometry and
the theory of $\ell $-adic sheaves. \

\subsubsection{Varieties}

In the sequel, we are going to translate back and forth between algebraic
varieties defined over the finite field $\mathbb{F}_{q}$ and their
corresponding sets of rational points. In order to prevent confusion between
the two, we use bold-face letters to denote a variety $\mathbf{X}$ and
normal letters $X$ to denote its corresponding set of rational points $X=%
\mathbf{X}(\mathbb{F}_{q})$. For us, a variety $\mathbf{X}$ over the finite
field is a quasi-projective algebraic variety, such that the defining
equations are given by homogeneous polynomials with coefficients in the
finite field $\mathbb{F}_{q}$. In this situation, there exists a (geometric)
\textit{Frobenius} endomorphism $Fr:\mathbf{X\rightarrow X}$, which is a
morphism of algebraic varieties. We denote by $X$ \ the set of points fixed
by $Fr$, i.e.,
\begin{eqnarray}
X=\mathbf{X}(\mathbb{F}_{q})=\mathbf{X}^{Fr}=\{x\in \mathbf{X}:Fr(x)=x\}.
\nonumber
\end{eqnarray}

The category of algebraic varieties over $\mathbb{F}_{q}$ will be denoted by
$\mathsf{Var}_{\mathbb{F}_{q}}$.

\subsubsection{Sheaves}

Let $\mathsf{D}^{b}(\mathbf{X)}$ denote the bounded derived category of
constructible $\ell $-adic sheaves on $\mathbf{X}$ \cite{BBD82, D81}. We
denote by $\mathsf{Perv}(\mathbf{X)}$ the Abelian category of perverse
sheaves on the variety $\mathbf{X}$, i.e., the heart with respect to the
autodual perverse t-structure in $\mathsf{D}^{b}(\mathbf{X})$. An object $%
\mathcal{F\in }\mathsf{D}^{b}(\mathbf{X)}$ is called $n$-perverse if $%
\mathcal{F[}n]\in \mathsf{Perv}(\mathbf{X)}$. Finally, we recall the notion
of a Weil structure (Frobenius structure) \cite{D81}. A Weil structure
associated to an object $\mathcal{F\in }\mathsf{D}^{b}(\mathbf{X)}$ is an
isomorphism%
\begin{eqnarray}
\theta :Fr^{\ast }\mathcal{F}{\longrightarrow }\mathcal{F}.
\nonumber
\end{eqnarray}

A pair $(\mathcal{F},\theta )$ is called a Weil object. By an abuse of
notation we often denote $\theta $ also by $Fr$. We choose once an
identification $\overline{%
%TCIMACRO{\U{211a} }%
%BeginExpansion
\mathbb{Q}
%EndExpansion
}_{\ell }\simeq
%TCIMACRO{\U{2102} }%
%BeginExpansion
\mathbb{C}
%EndExpansion
$, hence all sheaves are considered over the complex numbers.

\begin{remark}
All the results in this section make perfect sense over the field $\overline{%
%TCIMACRO{\U{211a} }%
%BeginExpansion
\mathbb{Q}
%EndExpansion
}_{\ell }$, in this respect the identification of $\overline{%
%TCIMACRO{\U{211a} }%
%BeginExpansion
\mathbb{Q}
%EndExpansion
}_{\ell }$ with $%
%TCIMACRO{\U{2102} }%
%BeginExpansion
\mathbb{C}
%EndExpansion
$ is redundant. The reason it is specified is in order to relate our
results with the standard constructions of the Weil representation \cite%
{Ge77, Ho73}.
\end{remark}

Given a Weil object $(\mathcal{F},Fr^{\ast }\mathcal{F\simeq F})$ one can
associate to it a function $f^{\mathcal{F}}:X\rightarrow
%TCIMACRO{\U{2102} }%
%BeginExpansion
\mathbb{C}
%EndExpansion
$ to $\mathcal{F}$ as follows
\begin{eqnarray}
f^{\mathcal{F}}(x)=\sum \limits_{i}(-1)^{i}\textrm{Tr}(Fr_{|H^{i}(\mathcal{F}_{x})}).
\nonumber
\end{eqnarray}

This procedure is called \textit{Grothendieck's sheaf-to-function
correspondence}. Another common notation for the function $f^{\mathcal{F}}$
is $\chi _{Fr}(\mathcal{F)}$, which is called the \textit{Euler
characteristic} of the sheaf $\mathcal{F}.$

\subsection{Canonical system of geometric intertwining kernels}

We shall now start the geometrization procedure.

\subsubsection{Replacing sets by varieties}

The first step we take is to replace all sets involved by their geometric
counterparts, i.e., algebraic varieties. The symplectic space $(V,\omega )$
is naturally identified as the set $V=\mathbf{V}(\mathbb{F}_{q})$, where $%
\mathbf{V}$ is a $2n$-dimensional symplectic vector space in $\mathsf{Var}_{%
\mathbb{F}_{q}}$. The Heisenberg group $H$ is naturally identified as the
set $H=\mathbf{H}(\mathbb{F}_{q})$, where $\mathbf{H}=\mathbf{V\times }%
\mathbb{A}^{1}$\ is the corresponding group variety. Finally, $\textrm{Lag}^{\circ }=%
\mathbf{Lag}^{\circ }(\mathbb{F}_{q})$, where $\mathbf{Lag}^{\circ }$ is the
variety of oriented Lagrangians in $\mathbf{V}$.

\subsubsection{Replacing functions by sheaves}

The second step is to replace functions by their sheaf-theoretic
counterparts \cite{Ga01}. The additive character $\psi :\mathbb{F}%
_{q}\longrightarrow
%TCIMACRO{\U{2102} }%
%BeginExpansion
\mathbb{C}
%EndExpansion
^{\times }$ is associated via the sheaf-to-function correspondence to the
Artin-Schreier sheaf $\mathcal{L}_{\psi }$ living on $\mathbb{A}^{1}$, i.e.,
we have $f^{\mathcal{L}_{\psi }}=\psi .$ The Legendre character $\sigma $ on
$\mathbb{F}_{q}^{\times }\simeq $ $\mathbb{G}_{m}(\mathbb{F}_{q})$ is
associated to the Kummer sheaf $\mathcal{L}_{\sigma }$ on $\mathbb{G}_{m}$.
The one-dimensional Gauss sum $G_{1}$ is associated with the Weil object
\begin{eqnarray}
\mathcal{G}_{1}\mathcal{=}\int \limits_{\mathbb{A}^{1}}\mathcal{L}_{\psi
(z^{2})}\in \mathsf{D}^{b}(\mathbf{pt}), \nonumber
\end{eqnarray}%
where, for the rest of these notes, $\int =\int_{!}$ denotes integration
with compact support \cite{BBD82}. Grothendieck's Lefschetz trace formula
\cite{Gr64} implies that, indeed, $f^{\mathcal{G}_{1}}=G_{1}.$ In fact,
there exists a quasi-isomorphism $\mathcal{G}_{1}{%
\longrightarrow }$ $H^{1}(\mathcal{G}_{1}\mathcal{)}[-1]$ and $\dim H^{1}(%
\mathcal{G}_{1}\mathcal{)=}1$, hence, $\mathcal{G}_{1}$ can be thought of as
a one-dimensional vector space, equipped with a Frobenius operator, sitting
at cohomological degree $1.$

Our main objective, in this section, is to construct a multiplicative system
of kernels $K:\textrm{Lag}^{\circ 2}\times H\longrightarrow
%TCIMACRO{\U{2102} }%
%BeginExpansion
\mathbb{C}
%EndExpansion
$ extending the subsystem $K_{U}$ (see \ref{sys_kernels_subsub}). The
extension appears as a direct consequence of the following geometrization
theorem:

\begin{theorem}[Geometric kernel sheaf]
\label{GIS_thm} There exists a geometrically irreducible $[\dim (\mathbf{Lag}%
^{\circ 2}\mathbf{)+n+1]}$-perverse Weil sheaf $\mathcal{K}$ on $\mathbf{Lag}%
^{\circ 2}\mathbf{\times H}$ of pure weight $w(\mathcal{K})=0$, satisfying
the following properties:

\begin{enumerate}
\item Multiplicativity property.\label{gmult_prop} There exists an
isomorphism
\begin{eqnarray}
\mathcal{K\simeq K\ast K}. \nonumber
\end{eqnarray}

\item Function property.\label{func_prop} We have $f_{|U}^{\mathcal{K}%
}=K_{U} $.
\end{enumerate}
\end{theorem}

For a proof, see Section \ref{GIS_sec}.

\subsubsection{Proof of Theorem \protect \ref{SSv-N2_thm}}

Let $K=f^{\mathcal{K}}$. Invoking Theorem \ref{GIS_thm}, we obtain that $K$
is multiplicative (Property \ref{gmult_prop}) and extends $K_{U}$ (Property %
\ref{func_prop}). Hence, we see that $K$ satisfies the conditions of Theorem %
\ref{SSv-N2_thm}. The nice thing about this construction is that it uses
geometry and, in particular, the notion of perverse extension which has no
counterpart in the set-function theoretic setting.

\section{Proof of the geometric kernel sheaf theorem \label{GIS_sec}}

Section \ref{GIS_sec} is devoted to sketching the proof of Theorem \ref{GIS_thm}%
.

\subsection{Construction}

The construction of the sheaf $\mathcal{K}$ is based on formula (\ref%
{ansatz2}).
Let $\mathbf{U\subset Lag}^{\circ 2}$ be the open subvariety consisting of
pairs $\left( M^{\circ },L^{\circ }\right) \in \mathbf{Lag}^{\circ 2}$ in
general position. The construction proceeds as follows:

\begin{itemize}
\item \textit{Non-normalized kernel.} On the variety $\mathbf{%
U\times H}$ define the sheaf%
\begin{eqnarray}
\widetilde{\mathcal{K}}_{\mathbf{U}}\left( M^{\circ },L^{\circ }\right)
=\left( \mathcal{\iota }^{-1}\right) ^{\ast }\mathcal{L}_{\psi }, \nonumber
\end{eqnarray}%
where $\iota =\iota _{M^{\circ },L^{\circ }}$ is the composition $\mathbf{%
Z\hookrightarrow H\twoheadrightarrow M\backslash H/L}$.

\item \textit{Normalization coefficient.} On the open subvariety $%
\mathbf{U\times H}$ define the sheaf%
\begin{equation}
\mathcal{C}\left( M^{\circ },L^{\circ }\right) \mathcal{=G}_{1}^{\otimes
n}\otimes \mathcal{L}_{\sigma }\left((-1)^{n\choose 2} \omega _{\wedge }\left(
o_{L},o_{M}\right) \right) [2n]\left( n\right) .  \label{norml_coef}
\end{equation}

\item \textit{Normalized kernels. } On the open subvariety $%
\mathbf{U\times H}$ define the sheaf%
\begin{eqnarray}
\mathcal{K}_{\mathbf{U}}=\mathcal{C\otimes }\widetilde{\mathcal{K}}_{\mathbf{%
U}}. \nonumber
\end{eqnarray}
\end{itemize}

Finally, take%
\begin{equation}
\mathcal{K=}j_{!\ast }\mathcal{K}_{\mathbf{U}},  \label{kernelsheaf}
\end{equation}%
where $j:\mathbf{U\times H\hookrightarrow Lag}^{\circ 2}\times \mathbf{H}$
is the open imbedding, and $j_{!\ast }$ is the functor of perverse
extension \cite{BBD82} (in our setting, $j_{!\ast }$ might better be called irreducible extension,
since the sheaves we consider are not perverse but perverse up to a
cohomological shift). It follows directly from the
construction that the sheaf $\  \mathcal{K}$ is irreducible $[\dim (\mathbf{Lag%
}^{\circ 2})+n+1]$-perverse of pure weight $0$.

The function property (Property \ref{func_prop}) is clear from the
construction. We are left to prove the multiplicativity property (Property %
\ref{mult_prop}).

\subsection{Proof of the multiplicativity property}

We need to show that%
\begin{equation}
p_{13}^{\ast }\mathcal{K\simeq }p_{12}^{\ast }\mathcal{K\ast }p_{23}^{\ast }%
\mathcal{K},  \label{gmult_eq}
\end{equation}%
where $p_{ij}:\mathbf{Lag}^{\circ 3}\times \mathbf{H}\rightarrow \mathbf{Lag}%
^{\circ 2}\times \mathbf{H}$ are the projectors on the $i,j$ copies of $%
\mathbf{Lag}^{\circ }$. We will need the following notations. Let $\mathbf{U}%
_{3}\subset \mathbf{Lag}^{\circ 3}$ denote the open subvariety consisting of
triples $\left( L_{1}^{\circ },L_{2}^{\circ },L_{3}^{\circ }\right) $ which
are in general position pairwisely.\ Let $n_{k}=\dim (\mathbf{Lag}^{\circ
k})+n+1$.

\begin{lemma}
\label{tech1_lemma} There exists, on $\mathbf{U}_{3}\times \mathbf{H}$, an
isomorphism
\begin{eqnarray}
p_{13}^{\ast }\mathcal{K\simeq }p_{12}^{\ast }\mathcal{K\ast }p_{23}^{\ast }%
\mathcal{K}. \nonumber
\end{eqnarray}
\end{lemma}

Let $\mathbf{V}_{3}\subset \mathbf{Lag}^{\circ 2}$ be the open subvariety
consisting of triples $\left( L_{1}^{\circ },L_{2}^{\circ },L_{3}^{\circ
}\right) \in \mathbf{Lag}^{\circ 4}$ such that $L_{1}^{\circ },L_{2}^{\circ
} $ and $L_{2}^{\circ },L_{3}^{\circ }$ are in general position. Lemma \ref%
{tech1_lemma} admits a slightly stronger form.

\begin{lemma}
\label{tech4_lemma}There exists, on $\mathbf{V}_{3}\times \mathbf{H}$, an
isomorphism
\begin{eqnarray}
p_{13}^{\ast }\mathcal{K\simeq }p_{12}^{\ast }\mathcal{K\ast }p_{23}^{\ast }%
\mathcal{K}. \nonumber
\end{eqnarray}
\end{lemma}

We can now finish the proof of (\ref{gmult_eq}). Lemma \ref{tech1_lemma}
implies that the sheaves $p_{13}^{\ast }\mathcal{K}$ and $p_{12}^{\ast }%
\mathcal{K\ast }p_{23}^{\ast }\mathcal{K}$ are isomorphic on the open
subvariety $\mathbf{U}_{3}\times \mathbf{H}$. The sheaf $p_{13}^{\ast }%
\mathcal{K}$ is irreducible $\left[ n_{3}\right] $-perverse as a pullback
by a smooth, surjective with connected fibers morphism, of an irreducible $%
[n_{2}]$-perverse sheaf on $\mathbf{Lag}^{\circ 2}\times \mathbf{H}$ .
Hence, it is enough to show that the sheaf $p_{12}^{\ast }\mathcal{K\ast }%
p_{23}^{\ast }\mathcal{K}$ irreducible $\left[ n_{3}\right] $-perverse.
Let $\mathbf{V}_{4}\subset \mathbf{Lag}^{\circ 4}$ be the open subvariety
consisting of quadruples 
$\left( L_{1}^{\circ },L_{2}^{\circ },L_{3}^{\circ },L_{4}^{\circ }\right) \in \mathbf{Lag}^{\circ 4}$ 
such that the pairs $%
L_{1}^{\circ },L_{2}^{\circ }$ and $L_{2}^{\circ },L_{3}^{\circ }$ are in
general position. Consider the projection $p_{134}:\mathbf{V}_{4}\times
\mathbf{H}\rightarrow \mathbf{Lag}^{\circ 3}\times \mathbf{H}$, it is
clearly smooth and surjective, with connected fibers. It is enough to show
that the pull-back $p_{134}^{\ast }\left( p_{12}^{\ast }\mathcal{K\ast }%
p_{23}^{\ast }\mathcal{K}\right) $ is irreducible $\left[ n_{4}\right] $%
-perverse. Using Lemma \ref{tech4_lemma} and also invoking some direct
diagram chasing one obtains%
\begin{equation}
p_{123}^{\ast }\left( p_{12}^{\ast }\mathcal{K\ast }p_{23}^{\ast }\mathcal{K}%
\right) \mathcal{\simeq }p_{12}^{\ast }\mathcal{K}\ast p_{23}^{\ast }%
\mathcal{K}\ast p_{34}^{\ast }\mathcal{K}.  \label{formula1}
\end{equation}

The right-hand side of (\ref{formula1}) is principally a subsequent
application of a properly normalized, Fourier transforms on $p_{34}^{\ast }%
\mathcal{K}$, hence by the Katz-Laumon theorem \cite{KL85} it is irreducible
$\left[ n_{4}\right] $-perverse.

Let us summarize. We showed that both sheaves $p_{13}^{\ast }\mathcal{K}$
and $p_{12}^{\ast }\mathcal{K\ast }p_{23}^{\ast }\mathcal{K}$ are
irreducible $[n_{3}]$-perverse and are isomorphic on an open subvariety.
This implies that they must be isomorphic. This concludes the proof of the
multiplicativity property.

%%%%%%%%%%%%%%%%%%%%%%%%%%%%%%%%%%%%%%%%%%%%%%%%%%%%%%%%%%%%%%%%%%%%%%
%%%%%%%%%%%%%%%%%%%%%%%% referenc.tex %%%%%%%%%%%%%%%%%%%%%%%%%%%%%%
% sample references
% "engineering"
%
% Use this file as a template for your own input.
%
%%%%%%%%%%%%%%%%%%%%%%%% Springer-Verlag %%%%%%%%%%%%%%%%%%%%%%%%%%

%
% BibTeX users please use
% \bibliographystyle{}
% \bibliography{}

%%%%%%%%%%%%%%%%%%%%%%%%%%%%%%%%%%%%%%%%%%%%%%%%%%%%%%%%%%%%%%%%%%%%%%

%\printindex
\end{document}